\documentclass[journal]{IEEEtran}

\usepackage{graphicx}
\usepackage{epsf}
\usepackage{amsmath, amsfonts}
\usepackage{latexsym}
\usepackage{subfigure}
\usepackage{multirow}
\usepackage{multicol}
\usepackage[table]{xcolor}
\usepackage[hidelinks]{hyperref}

\begin{document}

\title{\fontsize{22}{28}\selectfont Human EMF Exposure in Wearable Networks\\for Internet of Battlefield Things}

\author{
Imtiaz Nasim and Seungmo Kim

\thanks{I. Nasim and S. Kim are with Department of Electrical and Computer Engineering, Georgia Southern University in Statesboro, GA, USA (e-mail: \{in00206, seungmokim\}@georgiasouthern.edu).}
}

\maketitle

\begin{abstract}
Numerous antenna design approaches for wearable applications have been investigated in the literature. As on-body wearable communications become more ingrained in our daily activities, the necessity to investigate the impacts of these networks burgeons as a major requirement. In this study, we investigate the human electromagnetic field (EMF) exposure effect from on-body wearable devices at 2.4 GHz and 60 GHz, and compare the results to illustrate how the technology evolution to higher frequencies from wearable communications can impact our health. Our results suggest the average specific absorption rate (SAR) at 60 GHz can exceed the regulatory guidelines within a certain separation distance between a wearable device and the human skin surface. To the best of authors' knowledge, this is the first work that explicitly compares the human EMF exposure at different operating frequencies for on-body wearable communications, which provides a direct roadmap in design of wearable devices to be deployed in the Internet of Battlefield Things (IoBT).
\end{abstract}

\vspace{0.2 in}


\IEEEpeerreviewmaketitle

\section{Introduction}\label{sec_intro}
{\color{black}
By integrating smart devices on the battlefield, military strategists are tapping into the Internet of Things (IoT) to hone their tactics. The massive deployment of devices from wearables to autonomous vehicles and drones is expected to transform tomorrow's military battlefields into a large-scale Internet of Battlefield Things (IoBT) ecosystem. But in order to craft a strong deployment strategy, we have to juggle a number of complex variables. Among those variables, the human soldiers' safety under electromagnetic field (EMF) generated from wearable devices has often overlooked \cite{spectrum}. To overcome this challenge, this paper provides the first comprehensive analysis framework for analyzing the human EMF exposure problem in an IoBT system. The proposed framework will expedite the deployment of the IoBT by providing precise guidelines on how, when, and where to place the various smart devices in relation to enemy forces.
}

Over the last decade, on-body devices with higher data rates have attracted a massive number of wearable users in the global market. Numerous antenna designs are suggested in the literature to meet this demand for ever increasing users \cite{chahat_electronics_14}. Supporting high data rate for wearable application is challenging because of the bandwidth scarcity when users operate multiple devices simultaneously. More spectrum and less interference of higher frequency bands such as millimeter-wave (mmW) have the potential to address this problem \cite{venugopal_access_16}. However, antennas operating at high frequencies adopt more directional radiation pattern due to the excessive path loss in order to increase the data rate at the user end. Such directivity elevates the EMFs generated by the transmitter.

\vspace{0.1 in}
\section{Related Work}\label{sec_related}
\subsection{Concern on EMF Exposure}
Recently, there are strong warnings from scientists around the world on the harmful impacts of exposure to EMFs on human health in wireless communications systems adopting highly concentrated EMF energy \cite{5gappeal_sep17}. International agencies such as the U.S. Federal Communications Commission (FCC) \cite{fcc01} or the International Commission on Non-Ionizing Radiation Protection (ICNIRP) \cite{icnirp98} set the maximum radiation allowed to be introduced in the human body without causing any health concern. The FCC suggests power density (PD) as a metric measuring the human exposure to EMFs generated by devices operating at frequencies higher than 6 GHz \cite{wu15}, whereas a recent study suggested that the specific absorption rate (SAR) is more efficient to determine the health issues especially when devices are operating very close to the human body at very high frequencies \cite{rappaport15}. Such high frequencies enable high-gain directional antenna arrays with radiation energy focused in certain directions, which can yield increased power deposition in the main lobe points towards the human body \cite{shrivastava17}. Focusing on the extremely close proximity of a wearable to the human body, this work prioritizes SAR over PD to show the human EMF exposure from on-body communications devices. 

According to \cite{gao12}, the U.S. Food and Drug Administration (FDA) provided insufficient information to conclude that radio frequency (RF) emissions posed no threats. It also mentioned that some individual studies suggested possible effects of long-term exposure to high EMFs that can impose damage to biological tissues. The World Health Organization (WHO) and U.S. FDA indicate that the current academic understanding and scientific evidence do not show a clear mechanism how EMFs generate adverse health outcomes on a human body. Therefore, they go on to claim that additional research is warranted to address any gaps in knowledge. In fact, WHO's International Agency for Research on Cancer (IARC) has classified RF fields as possibly carcinogenic to humans \cite{WHO2011}.

\subsection{Current Safety Guidelines}
Considering all these aforementioned information, the guideline set by ICNIRP is 2 W/kg for 10-g SAR for near-field RF exposure for frequencies ranging from 10-10000 MHz, while FCC sets the guideline at 1.6 W/kg for 1-g SAR for frequencies from 0.1-6000 MHz \cite{wu15}.

\subsection{Mitigation of EMF Exposure}
Few prior studies in the literature paid attention to human EMF exposure in communications system at above 6 GHz frequency spectrum \cite{wu15}\cite{rappaport15}\cite{sambo15}\cite{colombi18}\cite{chahat2012}\cite{love16}. Propagation characteristics at different mmW bands and their thermal effects were investigated for discussion on health effects of EMF exposure in mmW radiation \cite{wu15}. Emission reduction scheme and models for SAR exposure constraints are studied in recent works \cite{sambo15}\cite{love16}. However, this paper analyzes the maximum possible exposure that a human soldier can experience from multiple wearable devices operating at different frequencies.

\begin{figure}[t]
\centering
\includegraphics[width = \linewidth]{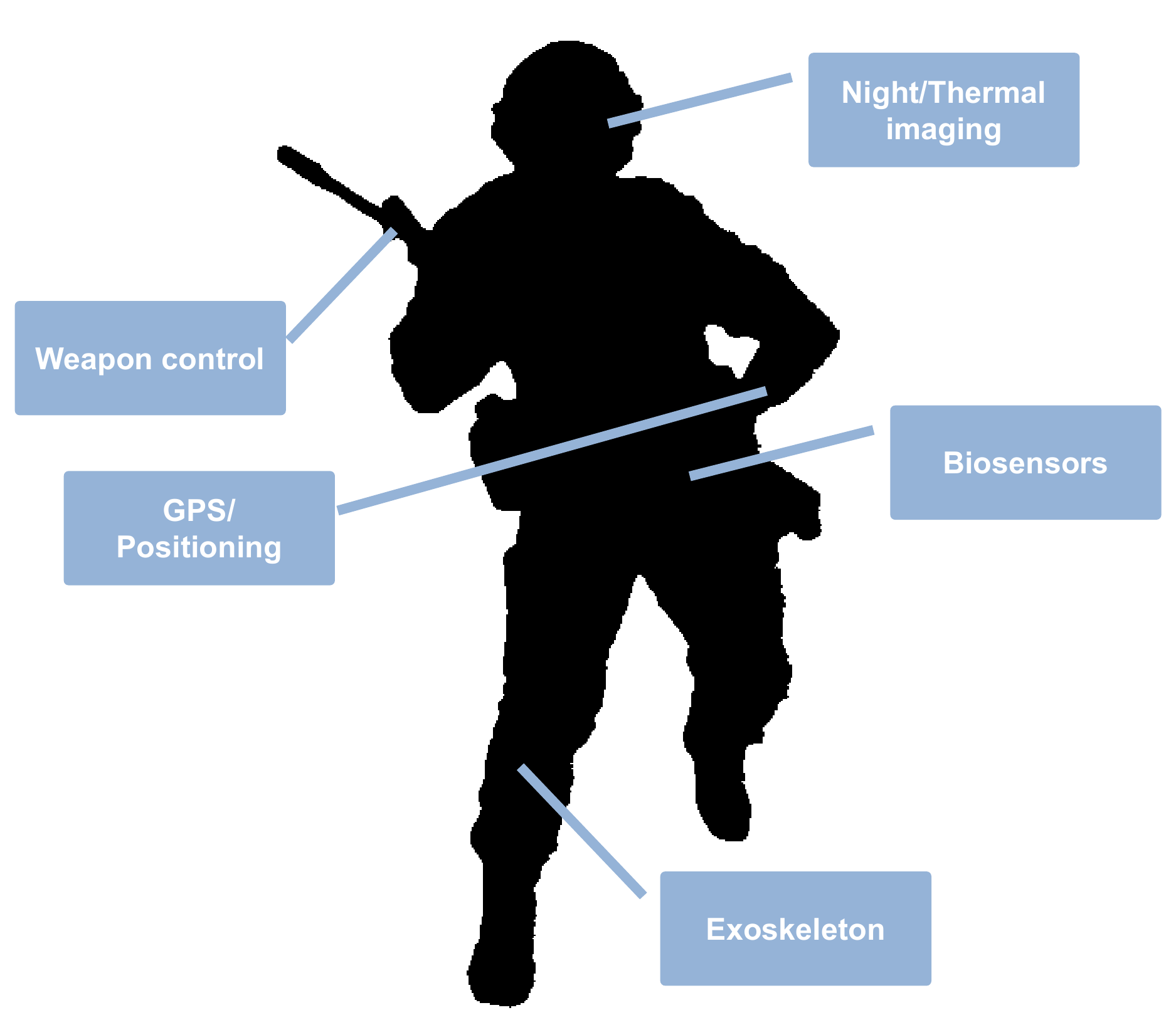}
\caption{An example of a soldier with multiple wearable devices \cite{alipour2010}-\cite{ibm}}
\label{soldier_example1}
\vspace{0.1 in}
\end{figure}

This present paper is distinguished from our previous studies in \cite{imtiaz_springer}\cite{imtiaz_southcon} in the sense that it addresses the human EMF exposure issues in the wearable communications environment, which has a more relevant insight for the military applications. Wearable devices can provide easier access to information and convenience for their users. They have varying form factors, from low-end health and fitness trackers to high-end virtual reality (VR) devices, augmented reality (AR) helmets, and smart watches \cite{sun2018}. Fig. \ref{soldier_example1} illustrates the possible nodes for wearable devices mounted on a soldier body. Most of the nodes are basically where a soldier gear is expected.

\subsection{Contributions of This Paper}
The contributions of this paper can be summarized as follows:
\begin{itemize}
\item It is the first work to the best of our knowledge that explicitly compares the human EMF exposure for on-body wearable communications at different frequencies. This work has a more relevant insight for military wearable applications.
\item It prioritizes SAR over PD to determine the health concerns for the wearable environment as the wearable network will generate an environment which is more like a near-field exposure.
\item It finds that the higher frequency application for the wearable environment with more directive pattern can elevate the human EMF exposure which has the potential to violate the existing regulatory guidelines.
\end{itemize}

\begin{table}[t]
\caption{Summary of key abbreviation and notation}
\centering
\begin{tabular}{|p{0.53 in} |p{2.5 in}|}
\hline
\textbf{Abbreviation} & \textbf{Description}\\
\hline
EMF & Electromagnetic field\\
IoBT & Internet of battlefield things\\
mmW & Millimeter wave\\
PD & Power density\\
SAR & Specific absorption rate\\
\hline\hline
\textbf{Notation} & \textbf{Description}\\
\hline
$G$ & Antenna gain\\
$\theta$ & Elevation angle\\
$\phi$ & Azimuth angle\\
$\delta$ & Depth of EMF penetration into human tissue\\
$f$ & Carrier frequency\\
\hline
\end{tabular}
\label{table_notation}
\end{table}

\section{System Model}\label{sec_model}
{\color{black}


Out of various spectrum bands, this paper focuses unlicensed bands, which enable cheaper and less complex devices (as well as longer battery life), all of which are desirable for wearable communications and networking \cite{sun2018}. The 2.4 GHz industrial, scientific, and medical (ISM) band has already long been coveted by numerous unlicensed communications systems. The FCC recently opened up an additional 7 GHz of spectrum available for unlicensed use through 64-71 GHz, which now provides a historic 14 GHz of contiguous chunk through 57-71 GHz--also known as the 60 GHz band \cite{jsac}.} Therefore, this paper considers a wearable communications environment with two different carrier frequencies--\textit{i.e.}, 2.4 GHz and 60 GHz. It enables a comparison of the impacts of carrier frequency on the human EMF exposure. Adopting two very different operating frequencies result in two different wearable environments. The adopted parameters for the two scenarios are summarized in Table \ref{table_parameters}.

It is noteworthy that we considered the maximum transmit power from on-body wearable devices with no power control nor adaptive radiation pattern adopted. The reason for such a worst-case assumption is to provide a `conservative' suggestion on human safety, which leaves some safety margin for the discussion of human EMF exposure compliance.

For simulation, we adopt the general antenna pattern equation \cite{seungmo17} given by
\begin{align}\label{eq_pattern_wifi}
G\left(\theta\right)=G_{max} - \exp\left(-2\pi j\delta \sin\theta\right) \rm{~[dB]}
\end{align}
where $\delta$ denotes the antenna element separation distance, and $\theta$ denotes a general angle. Notice that both azimuth and elevation angles are assumed to affect the antenna gain based on Eq. (\ref{eq_pattern_wifi}). The parameters $\theta_{3db}$ and $A_m$ for a wearable environment are considered as 93$^\circ$ \cite{cosan} and 30 dB, respectively.

It is more desirable to assume a continuous line-of-sight (LOS) \cite{heath_presentation_2015} link between a wearable device and the human body. Based on this rationale, this paper adopts a free space path loss (FSPL) model for calculation of a path loss, which is formally written as
\begin{align}\label{eq_fspl}
\text{PL} = 20\log(d) + 20\log(f) - 27.55 \rm{~[dB]}
\end{align}
where \textit{d} (m) is the distance between the antennas and \textit{f} (MHz) represents the operating frequency.

\begin{table}[t]
\small
\caption{Parameters for on-body wearable communications}
\centering
\begin{tabular}{|c|c|c|c|c}
\hline 
\textbf{Parameter}{\cellcolor{gray!10}} & 60 GHz {\cellcolor{gray!10}} & 2.4 GHz {\cellcolor{gray!10}}\\ \hline
\hline
Bandwidth & 2.16 GHz \cite{heath_presentation_2015} & 93 MHz \cite{cosan}\\
Max antenna gain & 11.9 dBi \cite{chahat2012} & 10.1 dBi\\
Transmit power & 10 dBm & 2 dBm \cite{wagih_2019}\\
Antenna elements & 16 \cite{heath_presentation_2015}  & 4\\
\hline
Path loss model & \multicolumn{2}{|c|}{Free space path loss (FSPL)}\\
Receiver gain & \multicolumn{2}{|c|}{10 dBi}\\
Wearable receiver noise figure & \multicolumn{2}{|c|}{6 dB}\\
Temperature & \multicolumn{2}{|c|}{290 K}\\ \hline
\end{tabular}
\label{table_parameters}
\end{table}

\subsection{Wearable antenna at 2.4 GHz}
\begin{figure}[t]
\centering
\includegraphics[width =.5\textwidth]{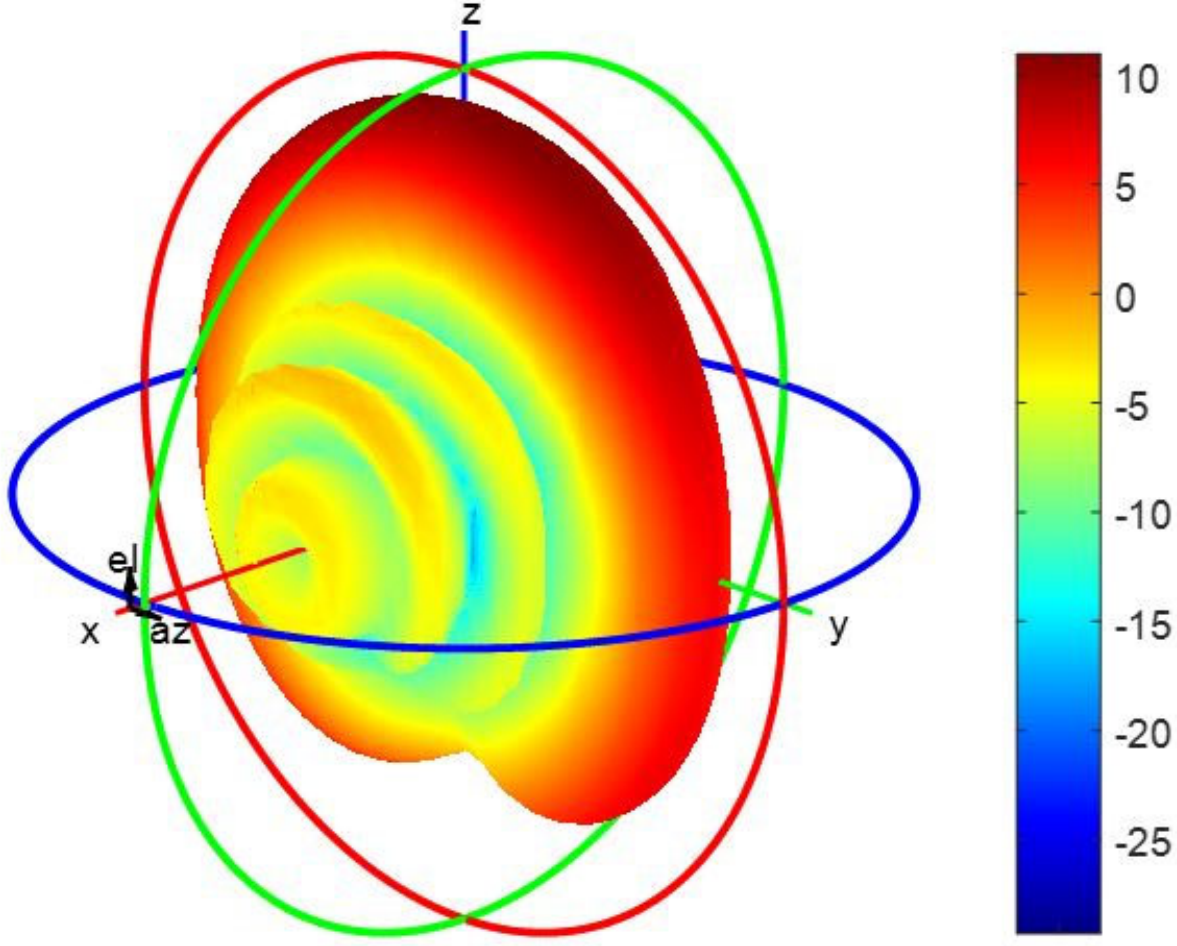}
\caption{Antenna pattern for wearable at 2.4 GHz}
\label{2_4pattern}
\end{figure}

Different communication links require different radiation patterns for the wearable operation. When two on-body devices communicate, an omnidirectional pattern seems to meet the requirements. However, when an on-body sensor communicates with an off-body device, a broadside pattern (or patch-like pattern) is more desirable. In order to support these two operating states, a patch like reconfigurable pattern is reported in \cite{yan2016} where the antenna can change its resonance to support both states with the same frequency. As such, this paper assumes a microstrip patch antenna which is more of omnidirectional in pattern for the operating frequency at 2.4 GHz. Fig. \ref{2_4pattern} depicts an omnidirectional antenna pattern operating at 2.4 GHz in a wearable environment.

\subsection{Wearable antenna at 60 GHz}
\begin{figure}[t]
\centering
\includegraphics[width =.5\textwidth]{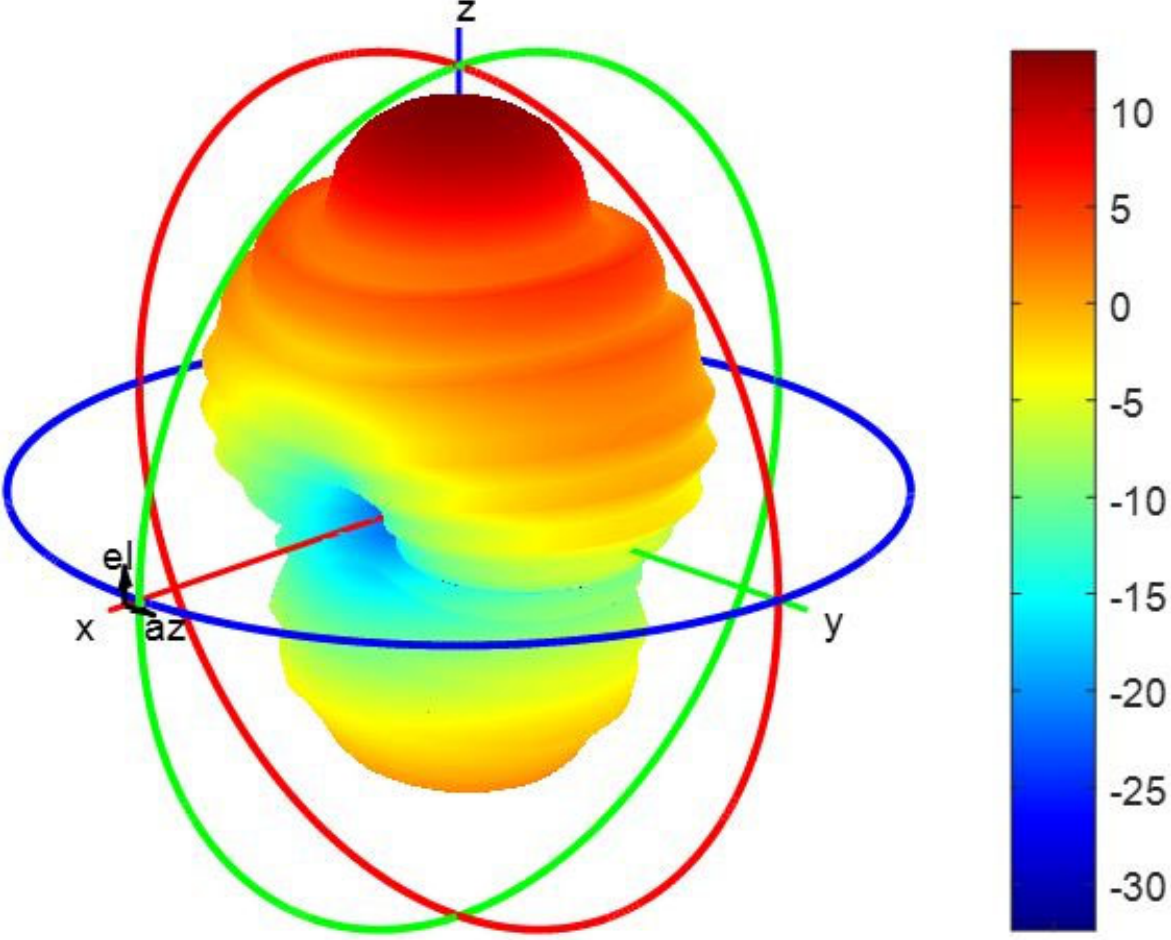}
\caption{Antenna pattern for wearable at 60 GHz}
\label{60_pattern}
\end{figure}

Commonly, wearable antennas are expected to be low-profile, lightweight, compact, and conformable to the body shape. In addition, because of high path loss at 60 GHz, medium gain values are desired. At 60 GHz, an end-fire pattern is required for which the direction of maximum radiation should be parallel to the body surface \cite{chahat2012}. A microstrip-fed Yagi-Uda antenna meets these specifications \cite{chahat2012} at 60 GHz. Fig. \ref{60_pattern} shows the antenna pattern of a Yagi-Uda antenna operating at 60 GHz which shows more of a directive radiation pattern compared to the one used at 2.4 GHz. We assumed 16 directors for the 60 GHz pattern as described in Table \ref{table_parameters}.

\vspace{0.2 in}
\section{Analysis of Human EMF Exposure}\label{sec_analysis}
The data rate from wearable devices can be calculated from Shannon's formula, which is given by
\begin{align}\label{eq_r}
R = B \log_2(1+\rm{SNR})
\end{align}
where $R$ and $B$ denote the data rate and bandwidth, respectively.

Biological effects of EMF depend on the level of energy absorbed into the human tissues. The depth of penetration into the human tissues depends on the frequency and conductivity of the tissues \cite{shrivastava17}. As mentioned in Section \ref{sec_intro}, above 6 GHz where many wearable systems will likely operate, safety guidelines \cite{fcc01}\cite{icnirp98} are defined in terms of PD due to the shallow penetration at such high frequencies. However, recent studies found that PD is not as useful as SAR or temperature in the assessment of EMF exposure since SAR can display the level of EMF energy that is actually `absorbed' in the body \cite{rappaport15}\cite{shrivastava17} while PD cannot. In general, PD cannot adequately evaluate the effect of human health impacts--\textit{e.g.}, temperature elevation of a direct contact area.

Moreover, some types of tissue--\textit{e.g.}, eyes--are more vulnerable to EMF-induced heating, which makes SAR and temperature to be considered as more adequate metrics in accurate measurement of human EMF exposure. Such differences achieved by the SAR becomes more critical in wearable applications since the wearable devices are mounted very close to the human body. In turn, it makes the scenario more relevant to the case of a near-field exposure. {\color{black}For this reason, this paper considers SAR as the primary metric for display of the human EMF exposure by a wearable device in IoBT}. Note that a higher SAR at the skin surface has the potential to increase the temperature at the human skin. Thus, our analysis indirectly gives an idea about the temperature elevation also for the wearable environment.

PD is defined as the amount of power radiated per unit volume at a distance \textit{d} \cite{wu15}, which is given by
\begin{align}\label{eq_pd_d}
\text{PD}\left(d\right) = \frac{\left|E\left(d\right)\right|^2}{\rho_0} \rm{~~[W/m^2]}
\end{align}
where $E\left(d\right)$ is the incident electric field's complex amplitude and $\rho_0$ is the characteristic impedance of free space. It can be rewritten by using the transmitter's parameters as
\begin{align}\label{eq_pd_phi}
\text{PD}\left(d, \theta, \phi\right) = \frac{P_T G_T\left(d, \theta, \phi\right)}{4 \pi d^2}
\end{align}
where $P_T$ is a transmit power; $G_T$ is a transmit antenna gain; $d$ is a antenna-human skin distance (m).

At high frequencies such as 60 GHz, most of the energy of a signal incident on human tissue is deposited into the thin surface of skin \cite{love16}. This can be expressed in terms of SAR, as a function of PD. SAR is defined as a measure of incident energy absorbed per a unit mass and a  unit time, which quantifies the rate at which the human body absorbs energy from an electromagnetic field. As such, it measures the electromagnetic energy that is dissipated per body mass. The local SAR value at a point $\mathbf{p}$ measured in W/kg \cite{love16} can be expressed as
\begin{align}\label{eq_sar_p}
\text{SAR}\left(\mathbf{p}\right) = \frac{\sigma\left|E\left(\mathbf{p}\right)\right|^2}{\rho} \rm{~~[W/kg]}
\end{align}
where $\sigma$ is the conductivity of the material and $\rho$ is the density of the material. The SAR at a point on the air-skin boundary \cite{chahat12} can be written as a function of PD$(d,\phi)$ as
\begin{align}\label{eq_sar_phi}
\text{SAR}\left(d,\theta, \phi\right) = \frac{2\text{PD}\left(d, \theta, \phi\right) \left(1 - \Gamma^2\right)}{\delta \rho}
\end{align}
where $\Gamma$ is the reflection coefficient \cite{wu15}; $\rho$ is the tissue mass density (1 $\text{g}/\text{cm}^3$ is used); and $\delta$ is the skin penetration depth (10$^{\text{-3}}$ m is used) \cite{rappaport15}.

Notice that the close proximity between a transmitter and the human tissue is the predominating factor in evaluation of human EMF exposure. As such, this paper displays the SAR as a function of distance, $d$, by averaging over all the possible angles to which an EMF is generated, which is formally written as
\begin{align}\label{eq_mean}
\overline{\text{SAR}}\left(d\right) = \frac{1}{(2\pi)^2} \int_{0}^{2\pi} \int_{0}^{2\pi} \text{SAR}\left(d,\theta, \phi\right) d\theta d\phi.
\end{align}
It is noteworthy that an angle can be widely varied as the person moves, since a wearable device is usually attached at a frequently moving body part--\textit{e.g.}, arm, leg, or head--as shown in Fig. \ref{soldier_example1}. It justifies such an `angle-averaged' quantity as $\overline{\text{SAR}}\left(d\right)$ the representative metric in display of the results.

\vspace{0.1 in}
\section{Simulation Results}\label{sec_results}
This section evaluates the simulation results of this work based on the EMF exposure analysis framework (described in Section \ref{sec_analysis}).

\subsection{Results}
\begin{figure}[t]
\centering
\includegraphics[width = \linewidth]{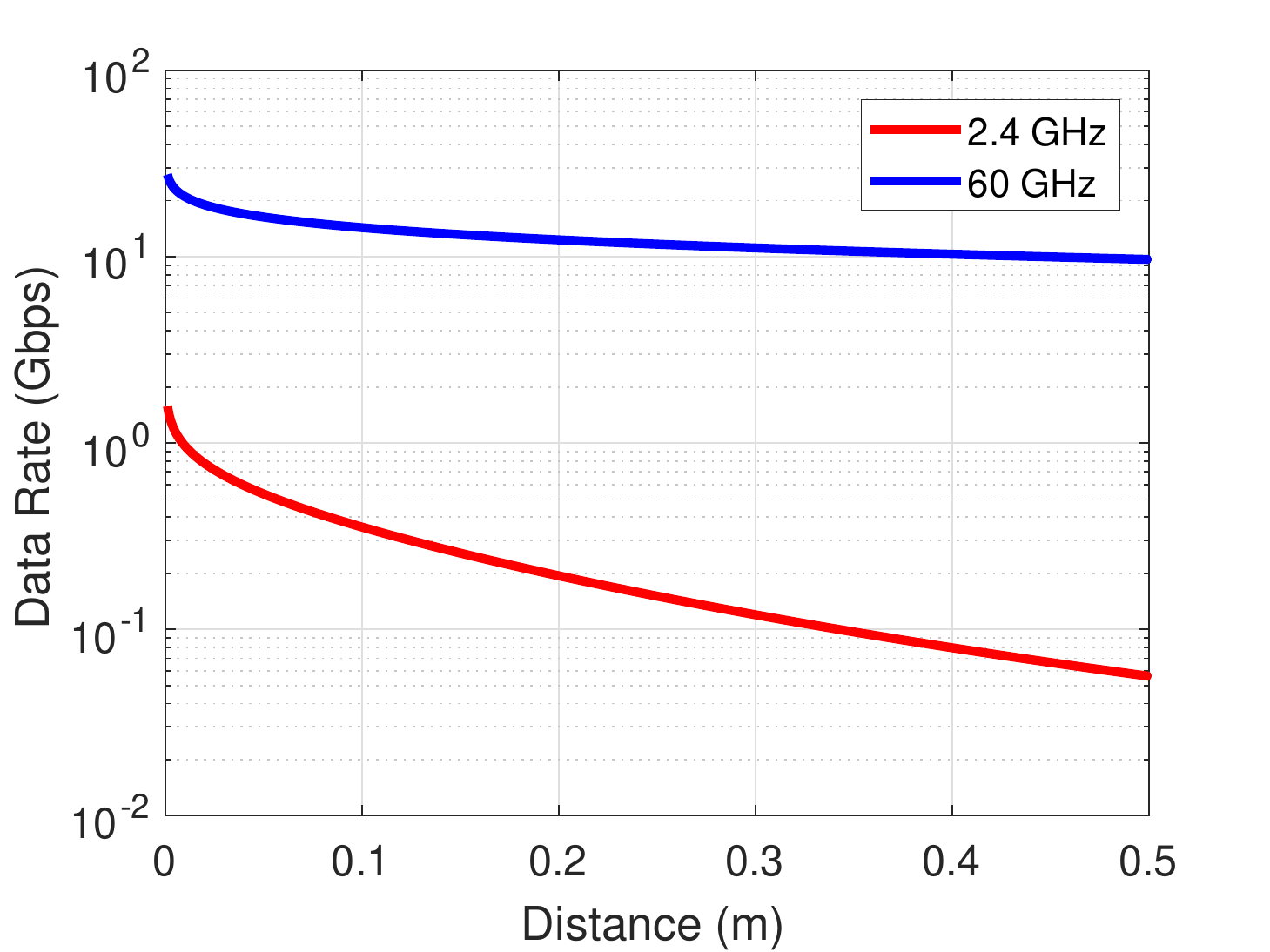}
\caption{Data rate versus distance}
\label{fig_rate}
\end{figure}

Fig. \ref{fig_rate} demonstrates the data rate that can be achieved from wearable devices operating at 2.4 GHz and 60 GHz considering the user's smart phones as the receiver \cite{moustafa2015}. It can be observed that devices operating at 60 GHz can provide remarkably higher data rates compared to 2.4 GHz. The rationale behind such an occurrence is the adaptation of directional antenna radiation pattern with a higher number of antenna elements at mmW frequencies as shown in Fig. \ref{60_pattern}. 

\begin{figure}[t]
\centering
\includegraphics[width = \linewidth]{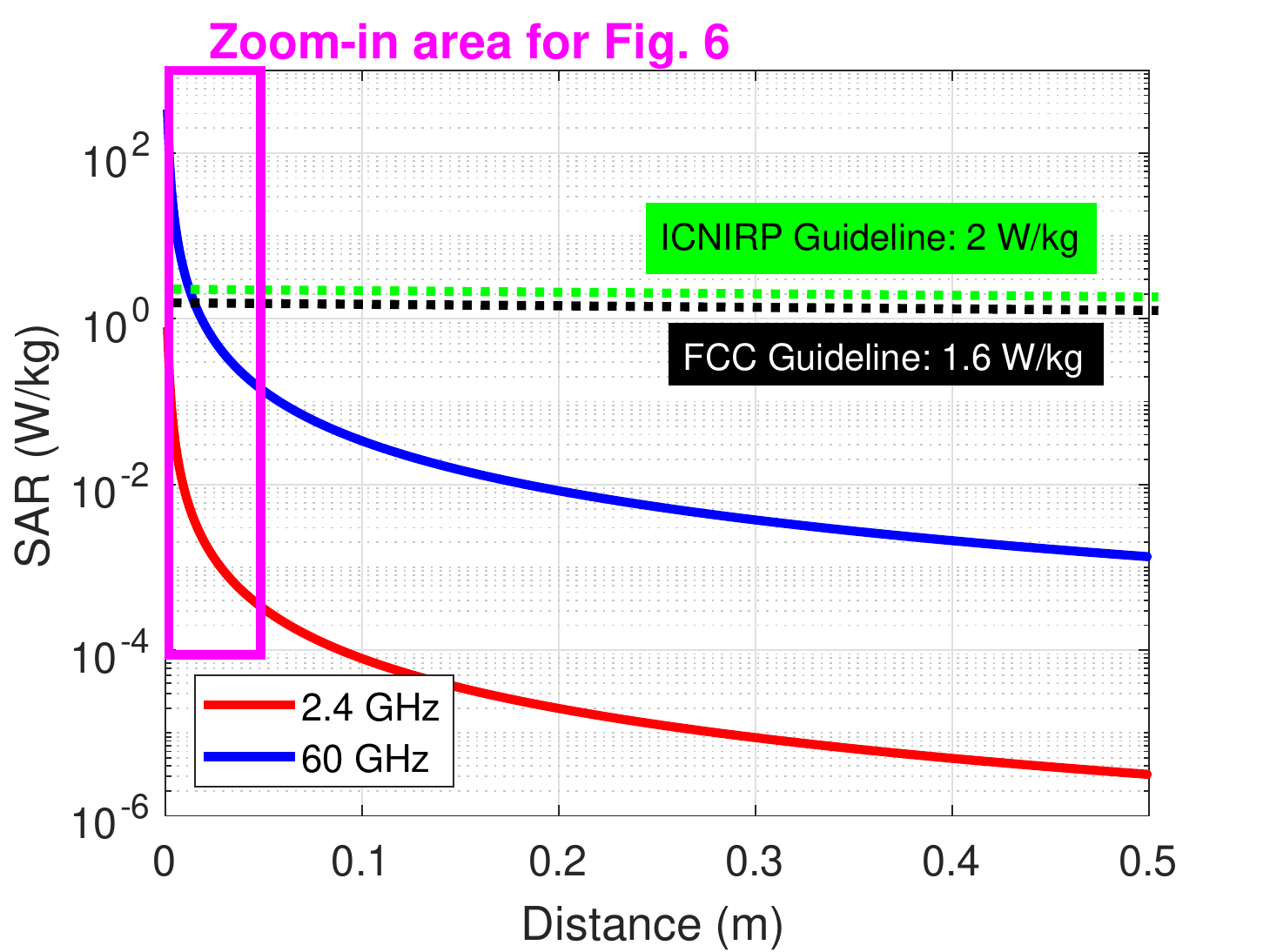}
\caption{SAR versus distance}
\label{fig_sar}
\end{figure}

Fig. \ref{fig_sar} represents the human EMF exposure from wearable devices in terms of SAR. The ICNIRP guideline for SAR near-field exposure is 2 W/kg for frequencies ranging from 10-10000 MHz for 10-g SAR. Our result in Fig. \ref{fig_sar} suggests that the average SAR measured at a certain distance from a wearable device can exceed the existing ICNIRP or FCC guidelines.

\begin{figure}[t]
\centering
\includegraphics[width = \linewidth]{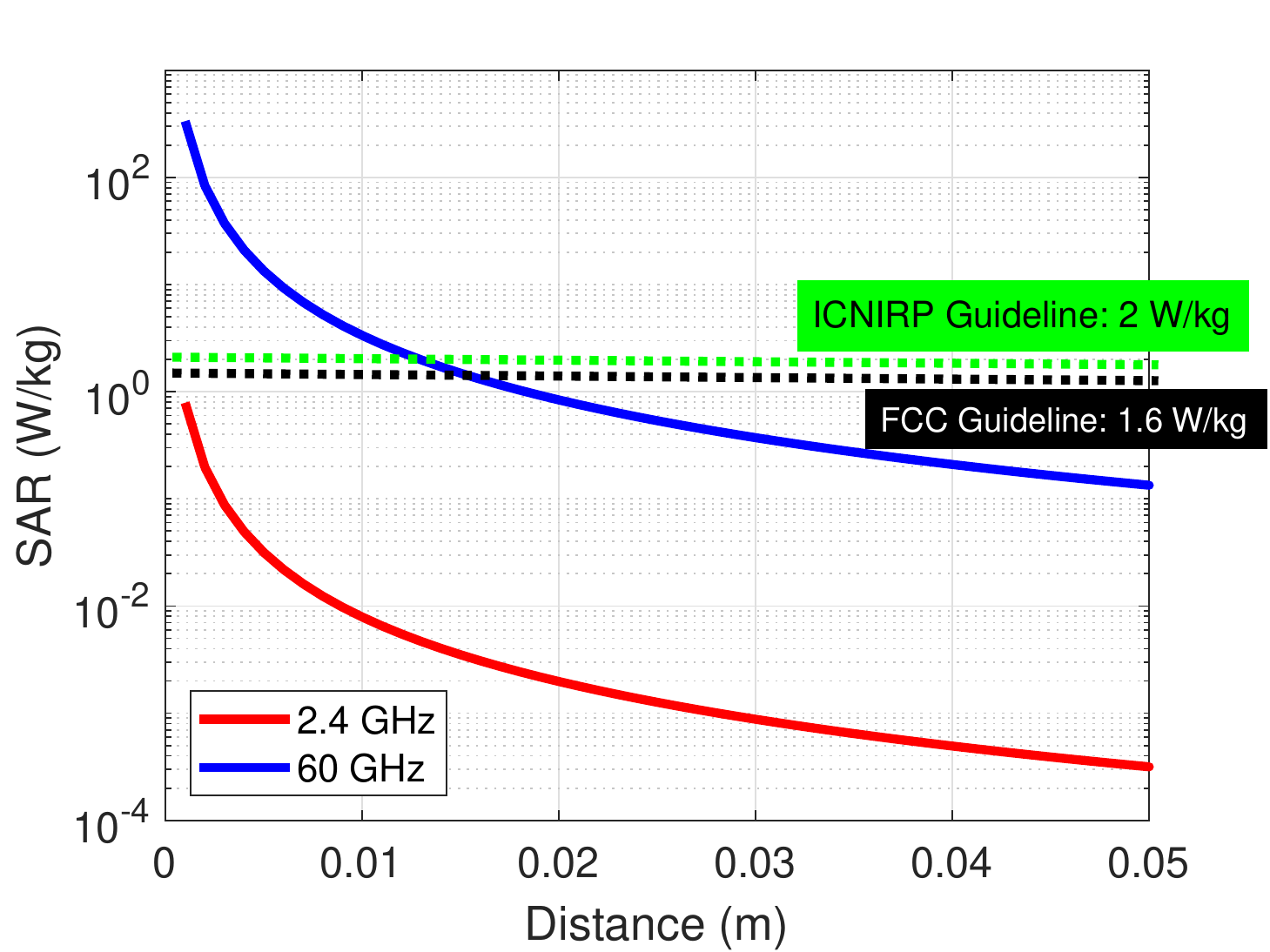}
\caption{SAR versus distance (a closer view within [0, 5] cm)}
\label{fig_sar_zoom}
\end{figure}

Fig. \ref{fig_sar_zoom} shows a zoomed-in display for the range of [0, 5] cm of Fig. \ref{fig_sar}. It provides a closer investigation of the minimum safe distance between the wearable devices that are mounted on the garments of the human body and the human skin. It can be suggested from Fig. \ref{fig_sar_zoom} that for a wearable device operating at 60 GHz, {\color{black} a separation distance of 12 and 15 mm is required according to the current ICNIRP and FCC guidelines, respectively}. However, the SAR for operating frequency at 2.4 GHz remains far below the guideline from the very first point.



\subsection{Discussions}
Although one can consider this distance to be very minimal, for a wearable environment, this small distance should also be taken into account for designing wearable devices on the human body. For instance, if a soldier is wearing an on-body device that is mounted on his smart glasses or the VR helmet, its radiation can impact sensitive organs like the eyes or the human brain. The adaptation of (i) higher number of antenna directors and (ii) higher transmit power are the primary reasons for the elevation of this human EMF exposure at 60 GHz. As noted previously, this elevation in SAR can increase the temperature at the surface of the human skin which may have a lethal impact on the human body when dosed in a continued manner or over a long-term period \cite{ziskin18}.

\subsection{Suggestions for Safe Deployment of Wearable Devices in IoBT}
Multiple key design insights for the IoBT communications were drawn from this paper's results. Specifically, they revealed the required separation distances between a wearable device and a human soldier's skin in terms of SAR. Furthermore, they showed the tradeoff with respect to carrier frequency between data rate and EMF exposure. A higher data rate could be achieved at 60 GHz thanks to a larger bandwidth; yet it caused a higher EMF absorption on the skin surface due to a shorter penetration depth.

As such, this work has various possible extensions. In this paper, we investigated the EMF exposure problem only within a single person's body. One possible extension is to incorporate discussions on impacts of EMF exposure between soldiers. For instance, there can be a battlefield scenario where a troop of soldiers wearing smart wearable devices operate with short distances from each other--\textit{e.g.}, search and attack or ambush \cite{adp12}. This can cause inter-person EMF exposure in addition to what was discussed in this paper. This extension may increase the complexity of analysis, considering numerous variables precisely such as the movement of each soldier.

\section{Conclusions}
This paper investigated human EMF exposure in a wearable communications applied for IoBT. Our simulation results showed that the human EMF exposure problem was highlighted when an operating frequency is used 60 GHz with more directive antenna radiation pattern. The safe distance considering the worst possible assumption was found as 12 mm at an operating frequency of 60 GHz with 16 directors placed in the wearable antenna. However, the higher operating frequency with a directive antenna pattern adopting higher number of antenna elements could increase the data rate up to 5 folds compared to the one at a relatively lower frequency such as 2.4 GHz. This work urges for more detailed investigation of human EMF exposure from on-body wearable devices, which will form solid foundational knowledge for safe operation of IoBT.


\end{document}